# The Self-Assembly of Nano-Objects Code: Applications to supramolecular organic monolayers adsorbed on metal surfaces

*Thomas Roussel[1,*] and Lourdes F. Vega[2]*

[1]Institut de Ciència de Materials de Barcelona, Consejo Superior de Investigaciones Científicas. (ICMAB-CSIC). Campus de la UAB, 08193 Bellaterra, Spain

[2]MATGAS Research Center (Carburos Metálicos/Air Products, CSIC, UAB). Campus de la UAB, 08193 Bellaterra, Spain

Abstract: The Self-Assembly of Nano-Objects (SANO) code we implemented demonstrates the ability to predict the molecular self-assembly of different structural motifs by tuning the molecular building blocks as well as the metallic substrate. It consists in a two-dimensional Grand Canonical Monte-Carlo (GCMC) approach developed to perform atomistic simulations of thousands of large organic molecules self-assembling on metal surfaces. Computing adsorption isotherms at room temperature and spanning over the characteristic sub-micrometric scales, we confront the robustness of the approach with three different well-known systems: ZnPcCl$_8$ on Ag$_{(111)}$, CuPcF$_{16}$ on Au$_{(111)}$ and PTBC on Ag$_{(111)}$. We retrieve respectively their square, oblique and hexagonal supramolecular tilling. The code incorporates generalized force fields to describe the molecular interactions, which provides transferability and versatility to many organic building blocks and metal surfaces.

1. INTRODUCTION

A predictive method to model the self-assembly of organic layers adsorbed on surfaces is essential to complete our understanding in tailoring new functional materials; their molecular nucleation, thermodynamic properties, phase equilibrium and growth mechanisms that occur at sub-micrometric scale. A deep control of the optimal conditions to precisely steer and monitor the first molecular building blocks is fundamental[1]. Metal-driven molecular architectronic[2] drastically depends on the first-deposited layer nanostructure. Plethora of possibilities allows the control and adjustment of the hetero-epitaxial film properties, as many as molecular building blocks (tectons) exist and choosing adequately the substrate structure. Scanning tunneling microscopy (STM) is widely used to elucidate the structural and electronic properties of the metal substrate and then the adsorbed organic layer. Basic principles behind molecular self-assembly of organic molecules on metal surfaces are pretty well understood[3] and their epitaxial organization on solid substrates[4-]. To a certain extent, quantum mechanics (QM) and density functional theory (DFT) approaches complement the experimental observation, from the electronic properties, understanding STM images[5], and also in determining the work function change of the new functional materials[6]. However, these simulations are limited to few hundreds of atom. On the other hand, atomistic classical molecular dynamics (MD) simulations can compute thousands of atoms providing insights on self-assembled ordered supramolecular layer. For instance, in the case of simulations of self-assembled shiff-based macrocyle molecules on gold surface[7] with an explicit solvent encompass $3.10^4$ atoms. However both approaches are computationally limited in system size and time scale if one aims to model the entire ordering process. Besides, large-scale modeling and statistical mechanics approaches are the key issue to understand and predict phase equilibrium. Indeed, supramolecular self-ordering typically occurs on large domains (nanometric scale) and at long time scale (minutes to hours)[8]. It implies modeling thousands of large molecules if one aims to describe phase transitions and the competitive interplay between the intermolecular and interfacial interactions. Monte Carlo simulation is a more suitable approach rather than Molecular dynamics (MD) simulations, which can not even simulate one second (in real-time) of self-assembly process. Reducing the physics of organic thin films to a two-dimensional system, different strategies can be found in the literature. Among them, kinetic Monte Carlo (kMC) simulation[9] is usually employed to describe the first steps of growth of organic molecular beam epitaxy deposition. Advantageously, it can predict phase diagrams of small molecules deposition. Assuming that the metallic substrate role was limited to constraining the molecular system in two dimensions, vertices in a specific symmetry of the substrate are predetermined. Combining empirical and this latter modeling approach, Silly *et al* have derived the hydrogen binding energies[10] as a function of the stoichiometric molecular mixture. Weber *et al* predict the structural stabilities diagram of open networks, compact phases, and high-temperature disordered phases, as a function of the temperature versus molecular coupling anisotropy[11]. Another interesting approach is to describe the molecular building block in a coarse-grained manner. For instance, Patti *et al* have recently represented amphiphilic cyclodextrins in its essential and most characteristic picture and reported the two-dimensional off-lattice MC simulations of their self-assembling behavior[12].

However, all the aforementioned approaches either loose the atomistic details of the molecules, which are represented in a simple shape and governed by defined simplified interaction rules, or assume lattices in an already defined symmetry, loosing the possibility of the organic overlayer to adjust its epitaxial properties regarding the surface. Another off-lattice strategies were recently proposed, combining MC/MD of rigid molecules taking advantage of the agent-based (AB) algorithm to study the self-assembly of a fully atomistic model of experimental interest[13,14,15]. It was shown that AB/MC gives a lower configuration energy in the same simulation time than both of the MC simulation techniques and the most conventional simulated annealing MC[16] method (commonly used for global energy minimization). However, the biased AB/MC technique does not respect the detailed balance.

Instead, we propose to tackle this issue performing Grand Canonical Monte-Carlo simulations (GCMC)[17], which minimize the configurational energy while it maximizes the entropy within fixed chemical potential and temperature. Still, atomistic simulations of thousands of large organic molecules require computational efforts that become rapidly unaffordable when the number of atom per molecule is large. However, assuming that the intramolecular degrees of freedom of the molecule and the substrate are playing a minor role on the self-ordering, one can freeze them. It allows the implementation of grid interpolation technique that reduces dramatically the computational time consumption. One commonly uses this approach to pre-calculate the interactions between an adsorbate with the adsorbent and performs GCMC calculations using grid interpolation technique, which in essence allows gas adsorption isotherms modeling[18-19] and also to determine zeolite templated carbon nanostructures[20]. Extending the grid interpolation technique to describe also the interaction between molecules, Mannsfeld *et al* have already demonstrated the need of computing thousands of organic large molecules[21-22] to simulate highly ordered organic thin films. Their approach aimed optimizing the hetero-epitaxial properties by simulated annealing MC[23], starting from the configuration observed experimentally. Ultimately, we propose a predictive numerical tool extending the latter method to the Grand ensemble in order to start from the bare surface and let the molecular building blocks self-order on their own.

Herein, we first present the results obtained on octachloro-zinc-phthalocyanines adsorbed on a dense silver surface ($ZnPcCl_8$ / $Ag_{(111)}$). The 2D-GCMC approach is highlighted by its ability to simulate thousands of molecules, necessary amount of molecules for reliable statistics. To illustrate the ordering behavior during the adsorption isotherms, we discuss the size distribution of the domains observed as a function of the chemical potential. Then, we present two other model systems studied to demonstrate the transferability of the approach. First, the adsorption of copper-phthalocyanines-fluorinated on gold surface ($CuPcF_{16}/Au_{(111)}$) for its oblique nanostructure[24] and interesting composition-dependent symmetry and crystallinity when co-adsorbed with Di-Indenoperylene (DIP) [25,26]. The second consists in the adsorption on silver of a five-fold ($C_5$) symmetry building block, the penta-*tert*-butylcorannulene ($PTBC/Ag_{(111)}$) [27]. Our interest for this latter molecule comes from its propensity to form long-range hexagonal order despite of its $C_5$ symmetry and for the diversity of phases observed experimentally when co-adsorbed with $ZnPcCl_8$.

2. SANO-GRID METHODOLOGY

2.1. Grid interpolation scheme:

Sub-micrometric scales simulations imply thousands of organic building blocks, which become computationally very demanding when the number of atom per molecule is large.

The grid-interpolation technique, first implemented by Mannsfeld *et al* [28-29], enables the calculation of both molecule-substrate and molecule-molecule interactions in a very fast way. It assumes the internal degrees of freedom frozen, for both the substrate and the molecular building blocks. Therefore, it allows pre-calculations of the potential energy between two building blocks and the building block with one unit cell of the surface.

The total potential energy of the system is expressed through two main contributions, the molecule-molecule interactions ($V^{MM}$) and molecule-substrate interaction ($V^{MS}$) as follows:

$$Etot = \sum_{I=1}^{Nmol} \sum_{J \neq I}^{Nmol} \left( V^{MM}(\theta_I, \theta_J, d_{IJ}) + V^{MS}(x_r^I, y_r^I, \theta_I) \right) \quad (1)$$

where $x_r^i$ and $y_r^i$ are the reduced coordinates of the molecular center of mass (COM) position (of the molecule indexed i) folded in the orthogonal unit cell of the metal surface; the intermolecular distance ($d_{ij}$) between the pairs of molecules *i* and *j* their respective COM, their reduced position and the two algebraic angles ($\theta_i$ and $\theta_j$) relatively to the axis **d**$_{ij}$, respectively extracted during the simulations from their two azimuthal molecular orientations ($\phi_i$ and $\phi_j$).

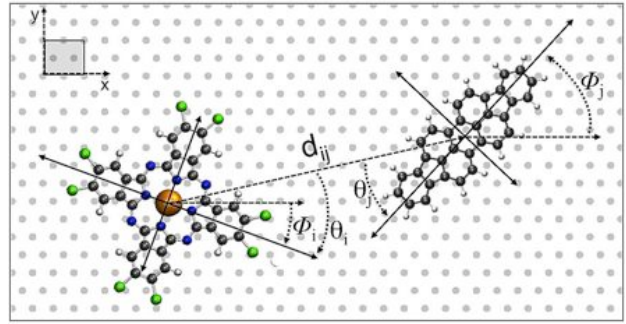

**Figure 1**. Scheme of the grid interpolation technique for the intermolecular potential energy between a ZnPcCl$_8$ and DIP.

Once the potential energy pre-calculation is achieved and stored, the computational time to evaluate the potential energy is not anymore function of the number of atom per molecule but only the number of molecules. Therefore, it reduces dramatically the computational time and allows performing full Monte-Carlo calculations. Then three preliminary steps are necessary before performing the Grand Canonical Monte-Carlo simulation: i) the optimization of the organic molecule atomic structure and the determination of the partial charge distribution generally provided by sophisticated QM and DFT approaches (*i.e.* Hyperchem[30] or Siesta[31] codes); ii) independently the metal surface structures (we propose in this work the aforementioned TB-SMA model to simulate large metal structures); iii) then the pre-calculation and storing of the molecule-surface and the intermolecular potential energies.

2.1.1. Pre-calculation of the interfacial interactions:

The interactions between the metal surface and the organic molecules are described atomically within a 12-6 Lennard-Jones pairwise potential (LJ). A recent set of parameters has been reported by Heinz *et al* [32] for several (*fcc*) transition metals (Ag, Al, Au, Cu, Ni, Pb, Pd, Pt). It reproduces accurately the quantitative and qualitative experimental surface tensions and interfacial properties of water molecules and organic molecules adsorbed on metal surfaces under ambient conditions. We pre-calculate the molecule-surface potential energy as a function of the reduced coordinates respectively to the surface unit-cell $(x_r^I, y_r^I)$, and $(\phi_I)$ the azimuthal molecular orientation. We mesh the orthorhombic surface unit cell with a spacing of 0.1 Å and 1 degree of molecular revolution. We store in a (2D+1)-grid the minimized potential energies after adjusting for each point the distance $(z_o)$ between the molecule COM and the surface. Hence, the potential energy is calculated atomically as:

$$V^{MS}\left(x_r^I, y_r^I, \phi_I\right)_{z_o} = \sum_{i=1}^{N_{atom}^{Molecule\ I}} \sum_{j=1}^{N_{atom}^{Metal}} \left(4\varepsilon_{ij}\left[\left(\frac{\sigma_{ij}}{r_{ij}}\right)^{12} - \left(\frac{\sigma_{ij}}{r_{ij}}\right)^{6}\right]\right)_{OPLS-AA}\bigg|_{z_o}$$

$$+ \left(\frac{1}{4\pi\varepsilon}\sum_{i=1}^{N_{CHARGES}}\left[\frac{-q_i^2}{2|z_i - z_M|} + \sum_{j\neq i}^{N_{CHARGES}} \frac{-q_i q_j}{\sqrt{\left(x_j - x_i\right)^2 + \left(y_j - y_i\right)^2 + \left(z_i + z_j - 2z_M\right)^2}}\right]\right)_{z_o} \quad (3)$$

where the first dispersive term (LJ) for the interatomic distances $r_{ij}$ between an the $i^{th}$ atom site of the molecule and the $j^{th}$ atom of the surface; The second term described in Mannsfeld's thesis[33] represents the Coulombic interactions between the partial charges localized on the atomic nuclei of the molecule and their respective mirror charges seen through free metal electrons considered as a *gelium*. This latter term is a function of the bi-dimensional coordinates of the molecule nuclei ($x_i$ , $y_i$) (parallel to the surface) where the partial charge is localized, and $z_i$ is the azimuthal distance when the molecule is distant at $z_o$ from the surface; $z_M$ designates the mirror plane position. Chulkov[34] reports them for several transition metals and for different crystalline surface orientations. The atomic parameters ($\sigma$ and $\varepsilon$) are taken from the OPLS All-Atom Force Field (OPLS-AA) parameters[35] for the organic molecules, and the ones reported by Heinz *et al* for the metal surfaces. The geometric mixing rule is applied to define the crossed-parameters. The SANO code is also implemented with CHARMM parameters[36], however, herein, we only discuss results obtained within the OPLS force field.

It is worthy to point out that the LJ parameters implicitly reproduce the essential features of the electronic metal structure, so that computed interfacial energies for flat surfaces are of the same accuracy as a tight binding or DFT methods, or quantum mechanical methods [37-38]. A major benefit is that computation times are about $10^6$ times shorter compared to *ab initio* methods. This latter important attribute allows the pre-calculation of the molecule-surface potential energy ($V^{MS}$) even if the surface has large unit cell, for instance when defects, vicinal surfaces, or reconstructed surfaces are considered.

2.1.2 Pre-calculation of the intermolecular interactions:

In this work, for a computational efficiency issue, the intermolecular weak interactions are described with a dispersive (LJ) and Coulombic terms; the latter is evaluated only in its real part being aware that the molecules are neutral and the methods will not diverge depending on the cutoff choice. The cutoff range is taken large enough (*e.g.* 3 to 5 nm in this work) to ensure 99% of its contribution between two molecules. The intermolecular interactions when performing Monte-Carlo simulations are then equivalent to the following atomistic calculation:

$$V^{MM}(\theta_I, \theta_J, d_{IJ}) = \sum_{i=1}^{\substack{N_{atom} \\ Molecule\ I}} \sum_{j=1}^{\substack{N'_{atom} \\ Molecule\ J}} \left( 4\varepsilon_{ij} \left[ \left(\frac{\sigma_{ij}}{r_{ij}}\right)^{12} - \left(\frac{\sigma_{ij}}{r_{ij}}\right)^{6} \right] \right)_{\text{OPLS-AA}} + \ldots \\ \sum_{i=1}^{\substack{N_{charges} \\ Molecule\ I}} \sum_{j=1}^{\substack{N'_{charges} \\ Molecule\ J}} \left( \frac{1}{4\pi\varepsilon} \left[ \frac{q_i\, q_j}{r_{ij}} \right] \right) \qquad (4)$$

We typically calculate (360x360x200) grid-points; the 360 orientations of both molecules, and 200 for the intermolecular distance. Note that the SANO code is also implemented for bi-molecular self-ordering. Therefore, three grids are needed to simulate the co-adsorption processes (A-A, A-B, B-B). It is worthy to point out here that the greater is the number of atoms per molecule, the longer is the pre-calculation. However, once the pre-calculation is done, the SANO code will compute with the same efficiency.

2.2. Generating a realistic transition-metal surface:

For this work, the metallic surface is modeled by a semi-empirical many-body potential derived from the tight binding scheme proposed by Ducastelle[39], in its second moment approximation (TB-SMA). The latter model from Rosato *et al*[40] implies that the band energy term of each atom is proportional to the square root of the second moment of the density of states, after that a pairwise Born-Mayer repulsive term is added to ensure core-repulsion.

The total energy of a system of N atoms is then written as:

$$E_{cohesion} = \sum_{i}^{N_{at}} \left( \sqrt{\sum_{j \neq i}^{N_{at}} \xi^2 \left( -\exp^{2q\left(\frac{r_{ij}}{r_o}-1\right)} \right)} + A \, \exp^{p\left(\frac{r_{ij}}{r_o}-1\right)} \right) \quad (2)$$

where $\xi$ in the attractive term is an effective hopping integral, summed over all the interatomic distances $r_{ij}$ (when $r_{ij} < r_c$) for a pair of atoms at sites *i* and *j*. The set of parameters (A, $\xi$, p, q) are adjusted on the bulk modulus, elastic constants and cohesive energies. Gupta[41] has reported that the based-on Friedel's thigh binding model yields physical contraction of face-centered-cubic (100), (110) and (111) surfaces, proper to the square root function of the atomic coordination number. Additionally, they have demonstrated that a simple pairwise potential yields intrinsically to a non-physical expansion of the interlayer distances at the metal surface. This latter unrealistic expansion would prejudice the adsorption potential energy calculation. Thus, one can render accurately bulk and surface properties, for instance for silver transition metal[42-43], and the interfacial properties of epitaxial silver clusters deposited on $MgO_{(100)}$ oxide[44]. Among others, we provide in the supplementary information (S.I.) different sets of parameters for the transition metals considered in this work and exhaustive references for other metal parameters. Note that a whole set of parameters has been reported recently to adjust bulk, elastic and surface energies properties at a time[45].

For this study, we generate the metal structures along the (111) orientation of the crystalline face-centered cubic (*fcc*) structure. We build slab structures made of 12 dense layers stacked along z-axis. We then perform quenched molecular dynamics[46-47] to optimize the interatomic distances for which the potential energy is minimized.

2.3 The 2D Grand-Canonical Monte-Carlo approach:

We have implemented a standard 2D-Grand Canonical Monte Carlo[48] (2D-GCMC) simulations, which advantageously insures the detailed balance[49]. Typically, we perform at least $10^6$ Monte-Carlo attempts per molecule to reach equilibrium. The attempts are either an insertion at a random place of a molecule, or a molecule is picked at random and then either extracted, translated, rotated, or translated and rotated at a time. The decision whether to translate or reorient is made at random and with an equal probability, as whether to insert or extract a molecule. In general, we perform 60% of canonical events and the 40% left for grand-canonical molecule exchange. The simulation box contains (n x m) orthorhombic unit cells of the substrate (*i.e*. $Ag_{(111)}$: 5.78 Å x 5.009 Å) and typically we will consider simulation boxes of (120x120) unit cells of the metal surface, meaning a simulation box of 693.6 Å per 601.08 Å for the silver surface (see figure 2). Periodic boundary conditions (PBC) are applied along the *x* and *y* directions. The simulations start from the bare surface, and we spread randomly few molecules on the substrate (*e.g.* less than 50). Adsorption isotherms allow the exploration of the accessible configurations space as a function of the incrementing chemical potential. Bruch *et al*. have reviewed the fundamental aspects of modeling physisorbed layers[50]. Interestingly, once a given coverage is reached, we can also perform simulations in the canonical ensemble (NVT) at a fixed number of molecules. Independently, one always set up the maximum displacement amplitude to half of the simulation box, moreover at low coverage, which insures an efficient exploration of the accessible configurations. Then, after $10^5$ Monte Carlo steps per molecule, the displacement amplitude is slowly calibrated to reach an average of 50% of successful canonical attempts, which insures an efficient equilibration[51].

3. CASE STUDIES:

Phthalocyanines (Pc) molecules have a large extended π-conjugated often implying flat-lying adsorbed molecules metal surfaces. These molecules are composed of 57 atoms and a metal is located in their center of mass, which can be substituted; by changing the nature of the metal, we can tune their optical (Zn, Cu), electronic (As, Ge) or magnetic (Fe, Co) properties. Fundamentally, these molecules are ideal systems since we can modify and enhance their adsorption potential energy on the substrate, and also their intermolecular interactions by functionalizing or substituting their peripheral atoms. This latter modification can imply substrate-imposed stress on the overlayer, which can have a significant influence on the resulting molecular self-assembly in some cases[52,53]. Therefore, our first study has been focusing on the adsorption of $ZnPcCl_8$ on $Ag_{(111)}$.

3.1. First case study: the adsorption of $ZnPcCl_8$ molecules on $Ag_{(111)}$:

The adsorption of octachloro-zinc-phthalocyanines molecules ($ZnPcCl_8$) on $Ag_{(111)}$ metal surface have been thoroughly reported in the last decade, by means of combined experimental and *ab initio* theoretical studies, showing compact arrangements governed by the activation of hydrogen–chlorine bonds networks[54]. Three different phases which evolve with a relatively long time (hours) were observed experimentally[55]; first, a rhombic (P1) phase where no hydrogen bonds are formed and presumed to be stabilized by permanent dipoles localized on the chlorines when they are facing; an asymmetric (P2) phases with four eight hydrogen–chlorine bonds per molecule, and finally the most stable (P3) square-phase stabilized by eight hydrogen–chlorine bonds. Experimentally, the lattice parameters *a* and *b* measured for each phase are the following: P1 (lattice parameters $a_1 = b_1 = 18$ Å), intermediate phase $P_2$ ($a_2 = 15$ Å, $b_2 = 18$ Å), and final phase P3 ($a_3 = 15$ Å, $b_3 = 15$ Å)[56]. Another subtlety of this system is the formation of stacking faults every three rows of molecules along the $[1\bar{1}0]$ in the P3 phase, which releases the stress cumulated from the incommensurate domain formed by the point-on-line coincidence with the substrate[57].

We start this study from the freestanding $ZnPcCl_8$ molecule structure reported by Oison[58] and its partial charges carried on each atom from DFT calculations. A detailed description of the nature of the hydrogen-chlorine contacts was also provided using first-principle DFT calculations. The figure bellow illustrates the typical self-ordered large domain of $ZnPcCl_8$ molecules on Ag$_{(111)}$, after performing GCMC calculation at T=400K and $\mu$=-910 kJ/mol, then relaxing the equilibrated configuration at 300K (NVT).

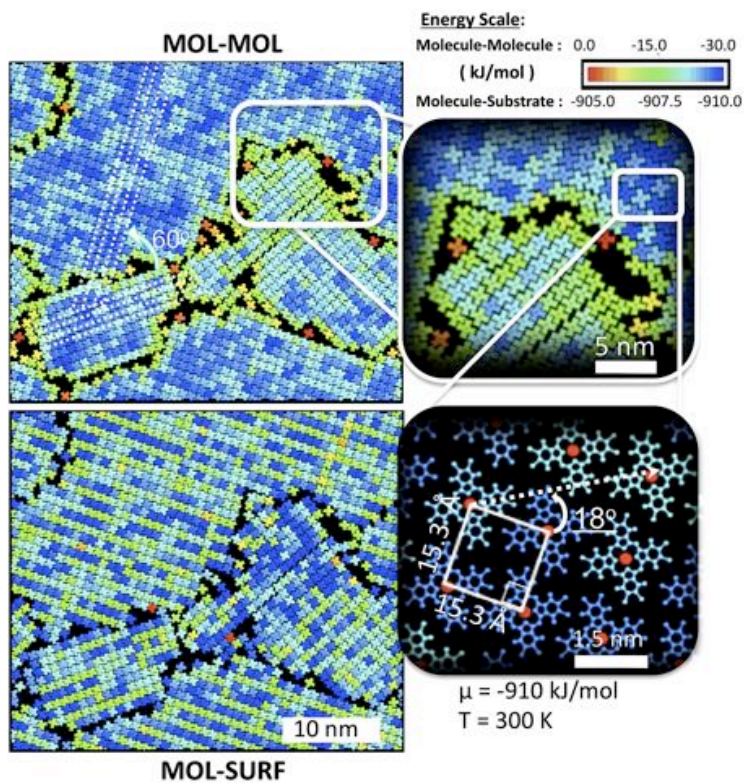

**Figure 2**. Snapshot of ZnPcCl$_8$ on Ag$_{(111)}$ after GCMC run at 400K and $\mu$ = -910 kJ/mol, then relaxed in NVT at 300K. The colors scales correspond to the molecule-molecule (*top left*) and molecule-surface (*bottom left*) potential energies (kJ/mol); the metal substrate is not represented. Insets represent zoomed region of the simulation box.

The configuration above shows around 1700 molecules adsorbed on the silver forming large domains disoriented and following two different dense orientations of the substrate. The configuration is represented twice, colored with the intermolecular and interfacial potential energies respectively (the bluer, the lower is the potential energy). The former tells about the defects, cluster contours, and the cumulated strain in the organic layer and the latter reveals directly the commensurability with the substrate. In agreement with the experimental observation and *DFT* calculations[59], it highlights the '*point-on-line*' coincidence[60] of the overlayer with the substrate. It means that the overlayer is orientated in a way that its two lattice vectors starts and ends on one class of the primitive substrate lattice lines. The closest matching super-cell of the silver surface is about 14.5 per 15 Å, respectively along the $[1\bar{1}0]$ directions. The averaged overlayer square-lattice spacing presents a higher mismatch with the former direction and thus cumulates tension along it. Note that the small clusters adjust their interfacial energy to the detriment of the intermolecular interactions, which can be also directly observed on the figure above. We perform adsorption isotherms at different temperature close to the ambient (Fig. 3).

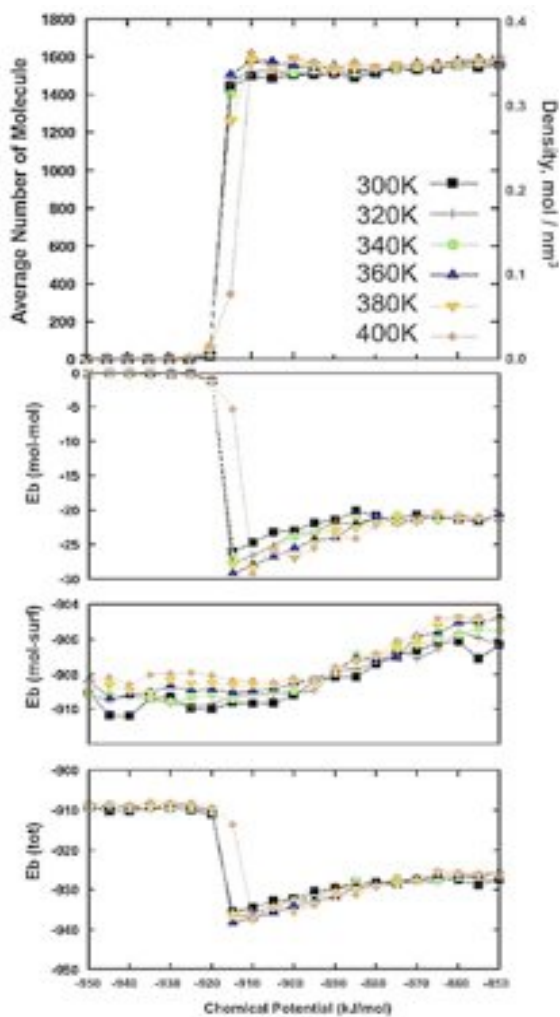

**Figure 3**. (*from top to bottom*): Adsorption isotherms of ZnPcCl$_8$ on Ag$_{(111)}$ as a function of the chemical potential (kJ.mol$^{-1}$) at six different temperatures: average number of molecules (*left axis*), and the corresponding molecular density (molecules.nm$^{-2}$); Follow the intermolecular, molecule-substrate and total potential energies (kJ.mol$^{-1}$).

We observe always first order transitions in the adsorption isotherms at chemical potentials corresponding to the molecule-substrate (~908 kJ/mol) and the potential energy of the first condensing cluster (from 5 to 12 kJ/mol). The effect of the temperature and the chemical potential is illustrated in the supplementary information (Fig. S1). Near the transition, large highly ordered P3 domains (*i.e.* several hundred of molecules per cluster) grow preferentially orientated relatively to the substrate. The simulation box is filled up to 1700 molecules, reaching molecular densities around 0.37 molecules per nm$^{-2}$. Note that the $P_3$ square phase has a maximum density of 0.39 molecules per nm$^{-2}$. The first order transition in the adsorption isotherms is accompanied with a steep descent in the total potential energy. Extracting the different contributions in the total energy as a function of the chemical potential, the intermolecular interactions drive the ordering. Also, when the temperature is raised of 100K, the loss observed in the molecule-substrate interaction is about 2 kJ.mol$^{-1}$ while the system gains about 5 kJ.mol$^{-1}$ from the intermolecular contribution. It is worthy to point out that the larger is the number of molecules the sharper is the first order transition and good are the statistics. An ultimate point within this preliminary work is that the intermolecular potential energy is around 20-30 kJ.mol$^{-1}$, which is of the order of magnitude binding energies reported for a square phase[61]. More importantly, the 2D-GCMC approach demonstrates its ability to optimize the intermolecular and interfacial properties right after the transition and appears as an efficient numerical tool to look for the most probable configurations at a given temperature.

We have implemented a numerical tool to analyze statistically the cluster size distribution as a function of the chemical potential. To illustrate it, we show the treatment of configurations at different temperature and represent the clusters in different colors to distinguish them and tag the clusters with indices corresponding to their number of molecule (see Fig. 4). The largest domains reach more than a thousand of molecules. The clusters were identified based on structural order criteria, in general, the distance and angle between first neighbors within a cutoff distance determined by the first g(r) peak (Fig. S2 in the S.I.). The average cluster sizes decay exponentially with the chemical potential. Near the transition, we encounter an average number of 600 molecules per cluster. When the chemical potential is far from the transition, the desorption acceptance rate starts to decay drastically and we observe small size clustering. Then, reporting the corresponding average number of cluster as a function of the chemical potential, it follows three distinguished trends. First, we observe a linear trend until no more than 15 large coexisting domains joined by either stacking defaults, or sharp grain boundaries. Still, most of the clusters managed to favor the intermolecular interactions and are guided by the $[1\bar{1}0]$ direction of the substrate. Then another linear trend with respect to the chemical potential corresponds to the clustering region of small size clusters with less than 100 molecules. This second regime corresponds to a blocked system with meta-stable jammed domains. Temperature annealing is then necessary to heal the organic layer. When the chemical potential is far above the transition, meaning a shift in chemical potential energy higher than the intermolecular interaction itself, the system does not even self-order. To avoid correlation between the cluster sizes and the simulation box size, a minimum of 1000 molecules has to be considered for these statistics.

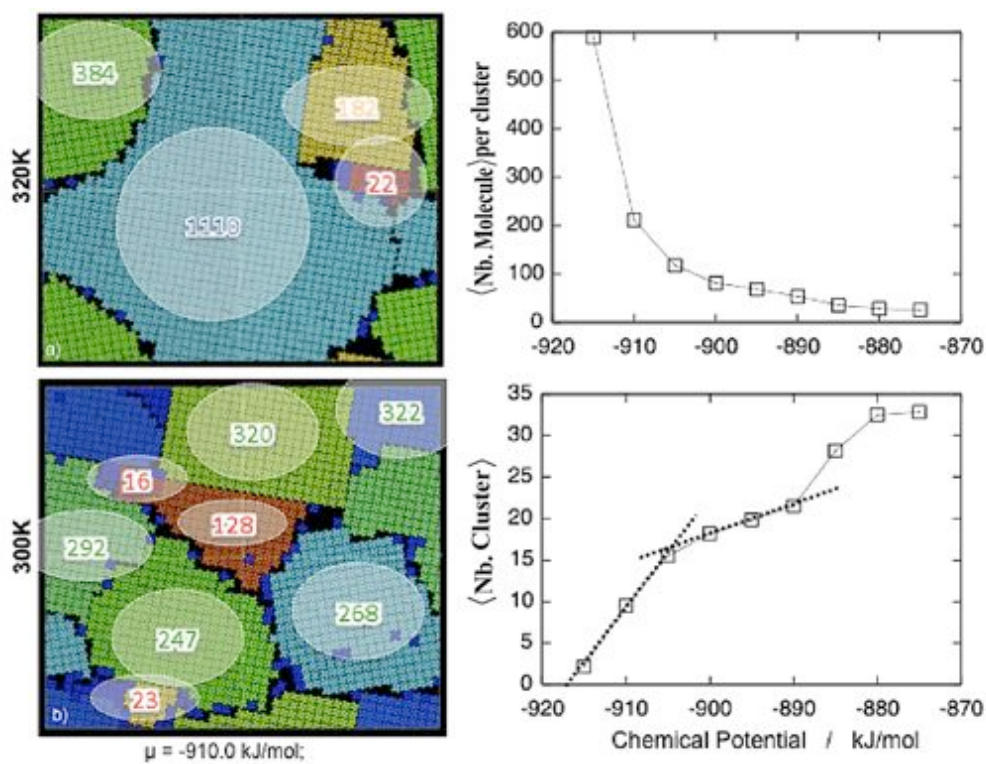

**Figure 4**. (*left*) The two structure snapshots of ZnPcCl$_8$ on Ag$_{(111)}$ at 320K (*top*) and 300K (*bottom*) at the same chemical potential $\mu$ = -910 kJ.mol$^{-1}$ are colored to distinguish the different clusters, and indices show their respective number of molecules. (*right*) The average number of molecules per cluster (*top*) and the average number of cluster are plotted as a function of the chemical potential (in kJ.mol$^{-1}$).

The annealing temperature effects and the incrementing chemical potential on the resulting nanostructures is illustrated on the supplementary information (Fig. S1). As a general trend, the higher is the chemical potential, or the lower is the temperature, the smaller are the cluster sizes with different alignments, more boundaries and defects. We can observe graphically the competitive alignment of each cluster along one of the three [1$\bar{1}$0] equivalent directions of the silver (111) surface.

Finally, the thermodynamic approach provides good qualitative and quantitative insights on the molecular self-ordering of ZnPcCl$_8$.

3.2. Method Transferability:

Myriad of research works have been arising in the last decade on supramolecular self-assembly of pi-conjugated organic molecules as building blocks to develop novel functional materials. Different means to control the symmetry and periodicity of the organic overlayer were proposed[62]. One can change the substrate lattice spacing or periodicity, for instance using a molecular template or surface reconstructions[63]. One can also tune the long-range ordering supramolecular nanostructure based on the electronic properties of the molecular building blocks. Notably for multi-component architecture[64], one can take advantage of the ability of the molecules to form stronger hydrogen-bonded networks, and tune the 2D physical properties to a wide range by adjusting the relative concentration. We choose two model systems and explore the transferability of the numerical approach and its capacity to model the self-assembly of molecular building blocks with different symmetry and on different surfaces. The first choice is the adsorption of fluorinated copper-phtalocynanine (CuPcF$_{16}$) on Au$_{(111)}$ for its known ability to form a dense oblique phase[65]. This latter molecule is of great interest when mixing it with the di-indenoperylene (DIP) molecule, which exhibit p-type and n-type conduction properties. The second choice is focused on the penta-*tert*-butylcorannulenes molecule (PTBC) adsorbed on Ag$_{(111)}$ for its ability to form hexagonal long-range ordered tessellations despite of its C$_5$ symmetry. Our interest for this molecule is also for its highly-ordered phases when co-adsorbed on Ag$_{(111)}$ with the aforementioned ZnPcCl$_8$ [66,67]. The following figure shows two configuration snapshots of $CuPcF_{16}$ molecules adsorbed on $Au_{(111)}$ and PTBC molecules adsorbed on $Ag_{(111)}$.

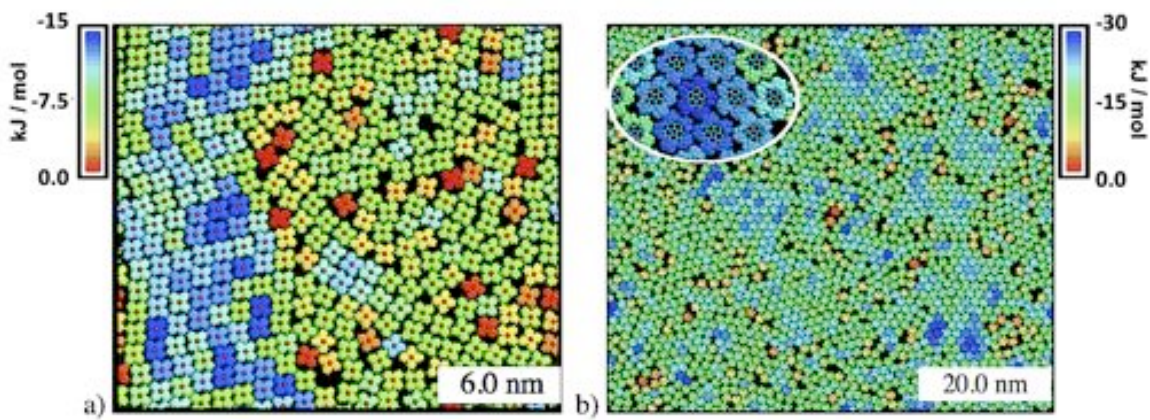

**Figure 5**. GCMC simulation at 300K of *a*) ~350 $CuPcF_{16}$ molecules adsorbed on $Au_{(111)}$ and b) ~2100 PTBC molecules on $Ag_{(111)}$. Colors are scaled to the intermolecular potential energies.

3.2.1. Adsorption of $CuPcF_{16}$ on $Au_{(111)}$:

Oblique packing was reported when fluorinated copper-phthalocyanines molecules are deposited on copper and gold (111) surfaces[68]. The atomic molecular structure of $CuPcF_{16}$ is extracted from DFT calculations[69]. We compute the GCMC simulations at 300K on a small simulation box of 30.6 x 26.5 $nm^2$. The maximum coverage reached is around 0.43 molecules/$nm^2$. The simulation leads to an oblique molecular tilling on gold surface. To illustrate the simulations behavior, we have represented a configuration where is observe a condensed phase co-existing with a disordered liquid like phase. The average lattice spacing (**a** = 14.7 ± 0.3; **b** = 15.2 ± 0.3 Å and γ = 73° ± 2°) is evaluated from the radial distribution function g(r), which is calculated with hundreds of configurations from the last equilibrium run (Fig. S3 in the S.I.). It is found in good agreement with the experimental arrangement already reported (a = 14.5 ± 0.8; b = 15.1 ± 0.8 Å and γ = 75º ± 2º). The alkaline interactions (F -- F) are weaker and less directional than the $ZnPcCl_8$ chlorine-hydrogen contacts. Also fluorine occupancy leaves less accessible room to the molecules to adopt a closer compact arrangement. It leads to the oblique phase instead of the square one observed with $ZnPcCl_8$. The intermolecular potential energy is about -14 kJ/mol, therefore much weaker than that for $ZnPcCl_8$. Interestingly, we observed few domains in a rhombic order co-existing with the oblique one. Also, the interactions with the substrate are still strong (-853 kJ/mol), however, weaker than the previously reported for $ZnPcCl_8$ adsorbed on silver. Adsorption on copper (111) surface has been also explored and similar supramolecular packing occurred, as reported experimentally.

### 3.2.2. Adsorption of PTBC on $Ag_{(111)}$:

Two-dimensional periodic tessellations of five-fold symmetry molecular building blocks have been reported [70-71] to be stabilized by phase transition blocking[72]. Calmettes *et al* have recently reported the hexagonal PTBC network adsorbed on silver $Ag_{(111)}$ [73] and the possibility to tune the symmetry of bimolecular stacking with $ZnPcCl_8$. They have highlighted a low interaction of the guest molecules with the substrate observing that PTBC networks were orientated independently of the Ag $[1\bar{1}0]$ direction, and also the need of low tunneling currents in order to avoid disordered molecule motions during the tip scan. We perform GCMC calculations at 300K on a simulation box of 69.41 x 60.11 nm$^2$ and we obtained about 2100 molecules at high coverage, meaning approximately a molecular density of 0.5 molecules/nm$^2$. We observe also large domains oriented independently of the $[1\bar{1}0]$ surface. The interfacial potential energy (-150 kJ/mol) is much weaker than that the metal phthalocyanines on silver and gold surfaces. Therefore, the intermolecular interactions (-27.5 kJ/mol) drive the hexagonal self-ordering. The lattice spacing of the close-compact hexagonal phase is extracted again from the g(r) (see figure S4 in C.I.) and is around 13.8 in average. Note that a shoulder in the first correlation peak arises at 13.3 Å corresponding to the most stable compact domains, in agreement with what has been measured experimentally at low temperature.

4. CONCLUSIONS

Thermodynamic properties and phase equilibrium of supramolecular organic layer adsorbed on surface are essential for a bottom-up device development. Typical molecular self-ordering occur on large domains, long time scale and imply hundreds to thousands of large molecules. The SANO code (Self-Assembly of Nano-Object) overcomes the computational limitation and allows large-scale modeling spanning over the characteristic sub-micrometric scales. During the molecular ordering, the method confers an intuitive understanding of the subtle and competitive interplay between intermolecular and interfacial interactions. Structural properties of three model systems are successfully reproduced, leading respectively to the square, oblique and hexagonal phases of $ZnPcCl_8$ adsorbed on $Ag_{(111)}$, $CuPcF_{16}$ on $Au_{(111)}$ and PTBC on $Au_{(111)}$.

Therefore, SANO code captures the essential physics of supramolecular self-assemblies. Taking benefits from the different generalized force fields, *i.e.* CHARMM and OPLS, the code is transferable and versatile. The main strength of the GCMC approach is to start the simulations from the bare surface. Therefore, one can perform readily numerical experiments starting only from the atomic structures of the molecular building blocks and the surface. Importantly, we treat thermodynamically the self-assembly at the experimental temperatures.

After this validation of the approach, the two-dimensional[‡] GCMC approach appears as a predictive numerical tool. The code is also extended to multi-component self-assembly. A large scope of physical properties becomes accessible as a function of the coverage. For instance the effect of the substrate roughness on the long-range molecular order; the cumulative stress on the organic layer due to the hetero-epitaxy; the commensurability for a given molecular superstructure; phase equilibrium of bi-molecular systems as a function of the relative concentration. Ultimately, the method is versatile and can be an interesting multi-scale approach if one aims to bridge quantum level calculations to the experimental scales and within a treatment in temperature.

ACKNOWLEDGMENT

T.R. is supported by a JAEdoc contract co-funded by CSIC and the European Union funds. T.R. thanks the Ministerio de Ciencia for financial support through projects FIS2009-13370-C02-02 and CONSOLIDER-NANOSELECT-CSD2007-00041. T.R. deeply acknowledges Dr. Jordi Faraudo for its helpful discussions. This work has been partially financed by the Spanish government, Ministerio de Ciencia e Innovacion, under projects CTQ2008-05370/PPQ. Additional support from Carburos Metalicos, Air Products Group, is also acknowledged.

FOOTNOTES

‡ the 3-dimensional version of SANO code is under development.

TABLE OF CONTENTS (TOC) GRAPHIC

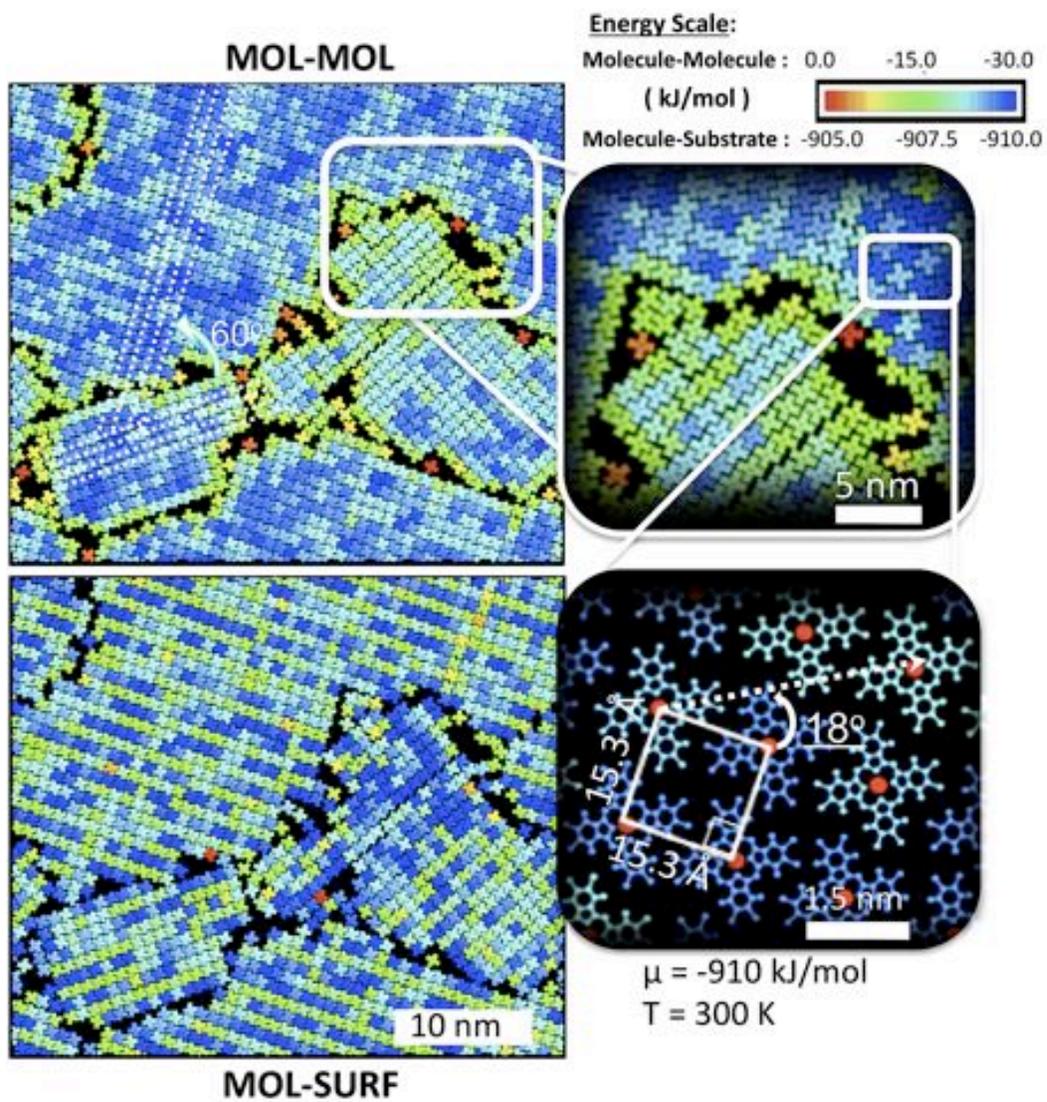

SUPPLEMENTARY INFORMATIONS:

The Self-Assembly of Nano-Objects Code: Applications to supramolecular organic monolayers adsorbed on metal surfaces

Thomas Roussel[1,*] and Lourdes F. Vega[2]

[1]Institut de Ciència de Materials de Barcelona, Consejo Superior de Investigaciones Científicas. (ICMAB-CSIC). Campus de la UAB, 08193 Bellaterra, Spain [2]MATGAS Research Center (Carburos Metálicos/Air Products, CSIC, UAB). Campus de la UAB, 08193 Bellaterra, Spain

Supplementary information provides: I) Annealing temperature, and chemical potential effects on the $ZnPcCl_8$ molecular ordering on $Ag_{(111)}$ surface. II) Structural analysis justifying the lattice spacing of the different structures ($ZnPcCl_8$ on $Ag_{(111)}$ ; $CuPcF_{16}$ on $Au_{(111)}$ ; PTBC on $Ag_{(111)}$) reported in the manuscript (radial distribution function and intermolecular angular distribution). III) Parameters implemented in the SANO code and taken from the following references.

To bring transferability to the SANO code, we have implemented OPLS-AA and CHARMM force fields; the former developed by Jorgensen *et al* [73] and the latter by MacKerell *et al* [73].

I: Annealing temperature, and chemical potential effects on the $ZnPcCl_8$ molecular ordering on $Ag_{(111)}$ surface:

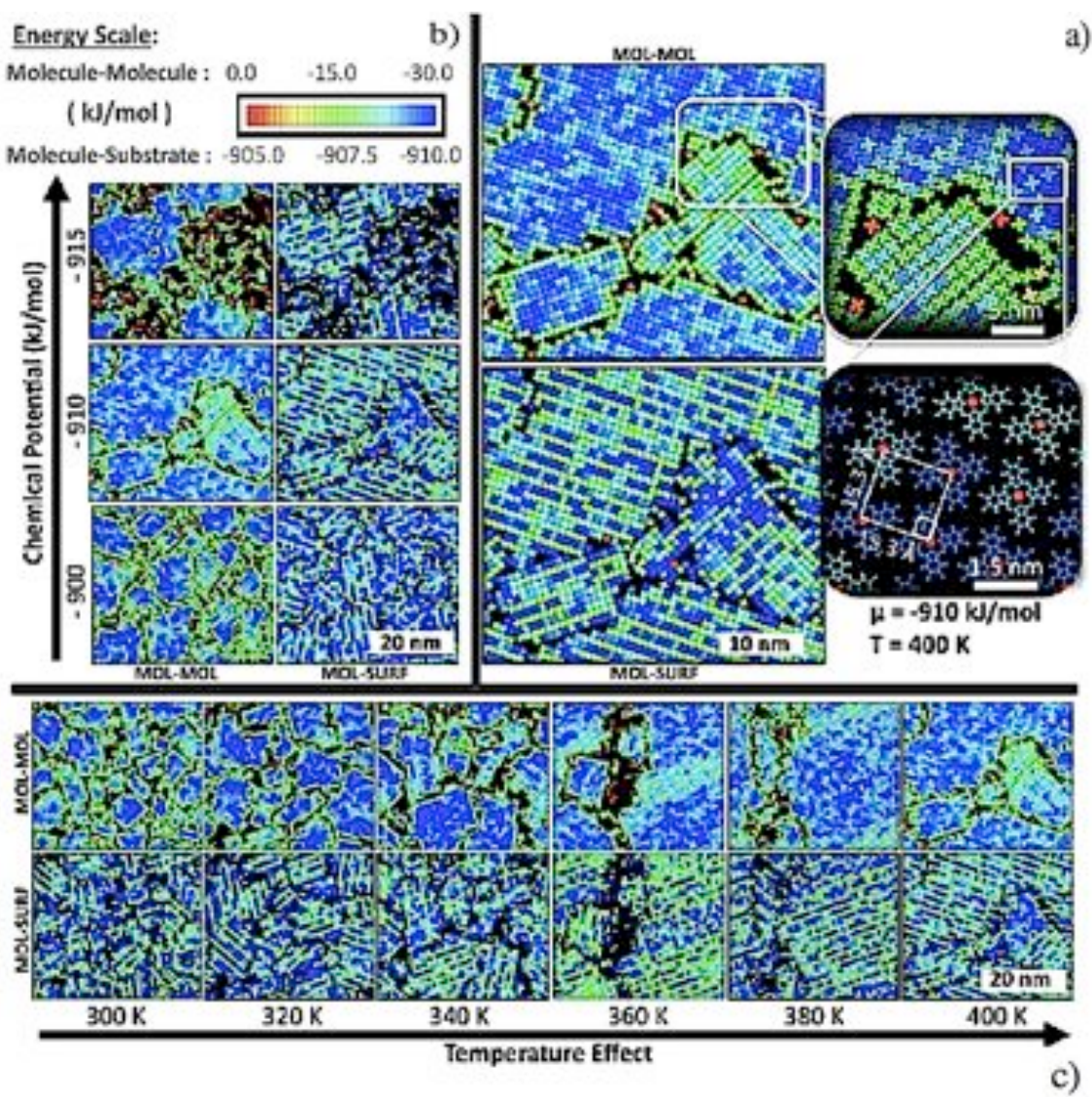

Figure S1: GCMC simulation of ZnPcCl$_8$ molecules self-assembled on Ag$_{(111)}$: a) structure snapshot at 400K $\mu$ = -910 (kJ.mol$^{-1}$). The colours are scaled on the potential energy (kJ.mol$^{-1}$) of the molecule-molecule (MOL-MOL) and molecule-surface (MOL-SURF); the substrate is not represented for visibility. Figure b) Effect of the chemical potential $\mu$ = -915, -910 and -900 (kJ.mol$^{-1}$), at 400K. Figure c) Annealing temperature effect on the cluster-size distribution near the transition.

II: Structural analysis:

II.1 - Adsorption at 300K of ZnPcCl$_8$ on Ag$_{(111)}$ :

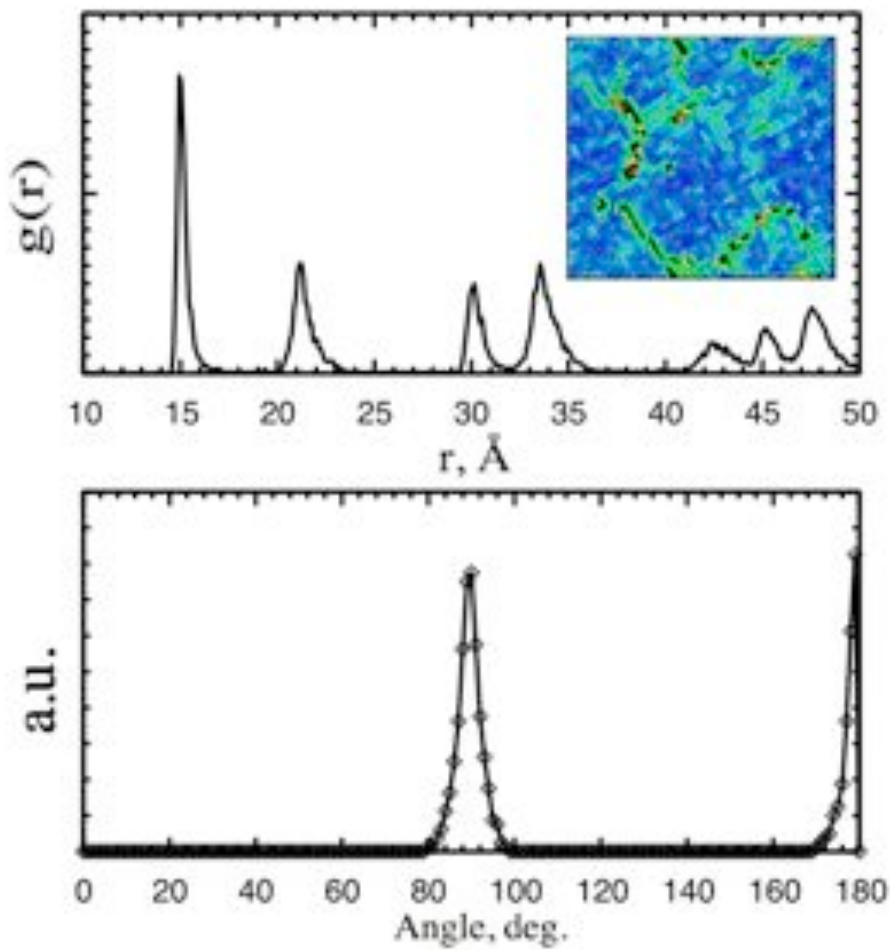

Figure S2: ZnPcCl$_8$ radial distribution function g(r) and intermolecular angular distribution between 1$^{st}$ neighbors with a cutoff distance evaluated from the g(r) first peak (statistics over 100 equilibrium configurations).

II.2 – Adsorption at 300K of CuPcF$_{16}$ on Au$_{(111)}$:

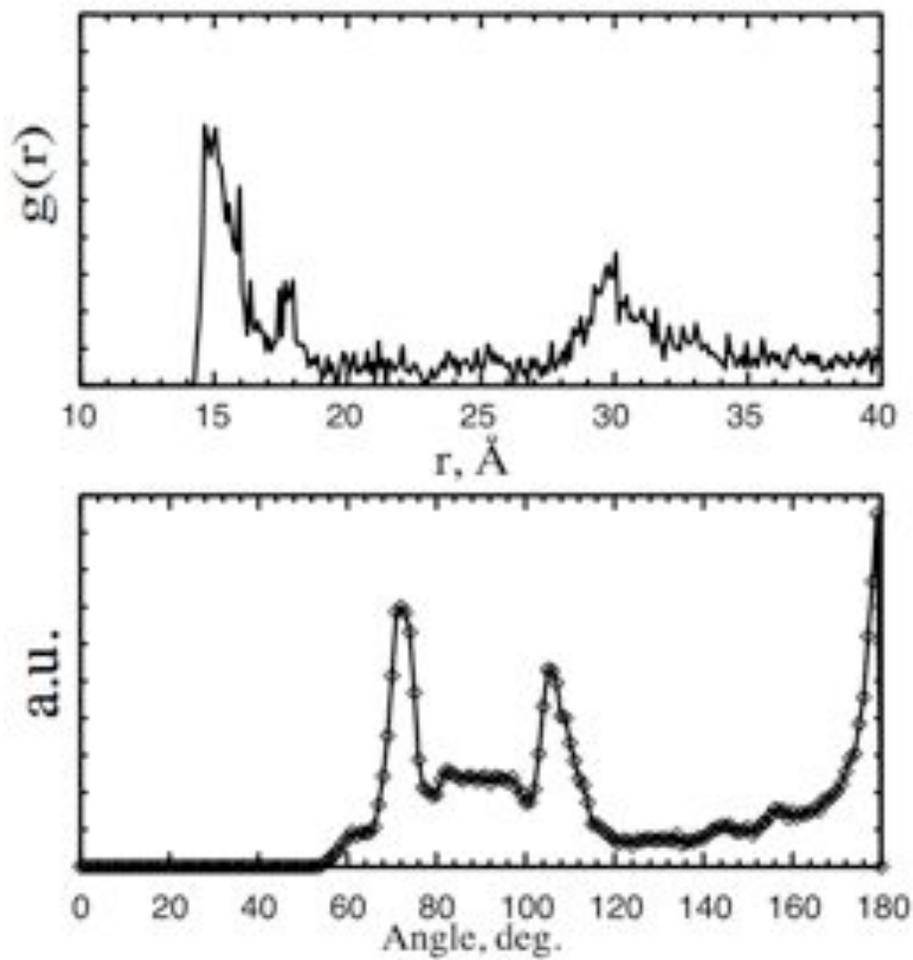

Figure S3: CuPcF16 radial distribution function g(r) and intermolecular angular distribution between 1$^{st}$ neighbors with a cutoff distance evaluated from the g(r) first peak (statistics over 100 equilibrium configurations).

II.3 - Adsorption at 300K of PTBC on Ag$_{(111)}$ :

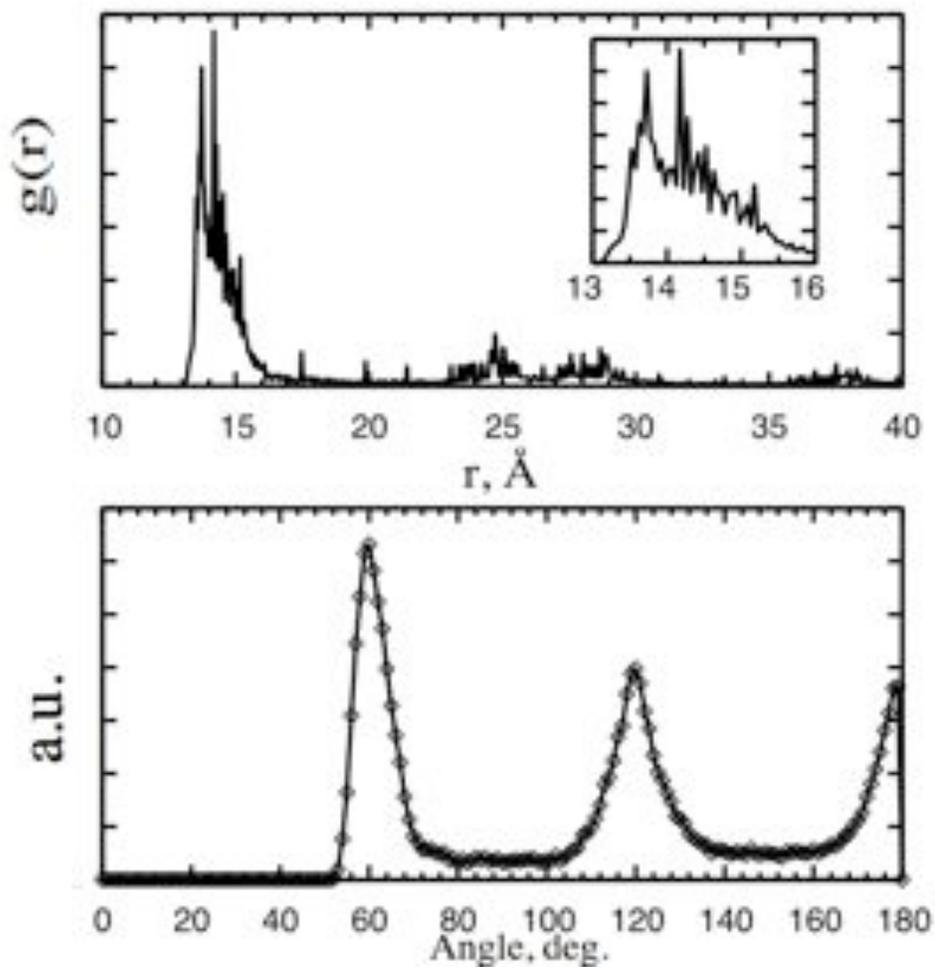

Figure S4: Penta-*tert*-butylcorannulenes (PTBC) radial distribution function g(r) and intermolecular angular distribution between 1st neighbors with a cutoff distance evaluated from the g(r) first peak (at 1statistics over 100 equilibrium configurations).

Table 1: Set of parameters of TB-SMA model used for the quenched molecular dynamics simulations to generate 12 layers-slabs of $Ag_{(111)}$, $Au_{(111)}$, and $Cu_{(111)}$. [73-73]

| TB-SMA | A (eV) | $\xi$ (eV) | P | q | $r_o$ (Å) | Bulk $E_{coh}$ (eV) |
|---|---|---|---|---|---|---|
| Ag | 0.10433 | 1.19402 | 10.790 | 3.19 | 4.09/√2 | 2.95 |
| Au | 0.209571 | 1.815276 | 10.139 | 4.033 | 4.08/√2 | 3.81 |
| Cu | 0.088530 | 1.274837 | 10.700 | 2.452 | 3.61/√2 | 3.49 |

Table 2: LJ Parameters for inter-atomic potential:

| | $\varepsilon_0 / k_B$ (kcal/mol) | $r_o$ (Å) |
|---|---|---|
| Ag | 4.56 | 2.955 |
| Au | 5.29 | 2.951 |
| Cu | 4.72 | 2.616 |

Note that those 12-6 and 9-6 LJ parameters were reported for several face-centered cubic metals (Ag, Al, Au, Cu, Ni, Pb, Pd, Pt) reproducing densities, surface tensions, interface properties with water and (bio)organic molecules, as well as mechanical properties in quantitative (<0.1%) to good qualitative (25%) agreement with experiment under ambient conditions. OPLS and CHARMM Lennard-Jones parameters for the metal surface are reported by Heinz *et al*. [73]

Table 3: Mirror plane distance from the metal surface reported by Chulkov *et al* [73] and implemented in SANO code:

|            | $Z_M$ (a.u.) |
|------------|--------------|
| $Au_{(111)}$ | 2.14         |
| $Ag_{(111)}$ | 2.22         |
| $Cu_{(111)}$ | 2.11         |

Note that Chulkov *et al* reviewed several image plane positions, from jellium calculation results for close-packed fcc, bcc and hcp structures for several transition metals. Those reported in table one are the one implemented in the beta-version of the SANO code. Results reported in the manuscript are using this image potential since molecules are close to the metal surface.